\documentclass{article}
\usepackage{spconf,amsmath,graphicx}
\usepackage{amssymb}
\usepackage{subfig}
\usepackage[hidelinks]{hyperref}
\usepackage{color}
\usepackage{siunitx}
\usepackage{booktabs}
\usepackage{etoolbox}
\usepackage{multirow}
\usepackage{amsmath}
\usepackage{changes}

\newrobustcmd{\B}{\bfseries}

\title{SASA: saliency-aware self-adaptive snapshot compressive imaging}
\name{ Yaping Zhao$^{1,2}$
 , Edmund Y. Lam$^{1,2,*}$}

\address{
$^1$ Department of Electrical and Electronic Engineering, The University of Hong Kong, Hong Kong SAR\\
$^2$ACCESS –- AI Chip Center for Emerging Smart Systems, Hong Kong SAR}
	
\begin{document}
\newcommand{\Amat}{{\bf A}}
\newcommand{\Bmat}{{\bf B}}
\newcommand{\Cmat}{{\bf C}}
\newcommand{\Dmat}{{\bf D}}
\newcommand{\Emat}[0]{{{\bf E}}}
\newcommand{\Fmat}[0]{{{\bf F}}}
\newcommand{\Gmat}[0]{{{\bf G}}}
\newcommand{\Hmat}[0]{{{\bf H}}}
\newcommand{\Imat}{{\bf I}}
\newcommand{\Jmat}[0]{{{\bf J}}}
\newcommand{\Kmat}[0]{{{\bf K}}}
\newcommand{\Lmat}[0]{{{\bf L}}}
\newcommand{\Mmat}[0]{{{\bf M}}}
\newcommand{\Nmat}[0]{{{\bf N}}}
\newcommand{\Omat}[0]{{{\bf O}}}
\newcommand{\Pmat}[0]{{{\bf P}}}
\newcommand{\Qmat}[0]{{{\bf Q}}}
\newcommand{\Rmat}[0]{{{\bf R}}}
\newcommand{\Smat}[0]{{{\bf S}}}
\newcommand{\Tmat}[0]{{{\bf T}}}
\newcommand{\Umat}{{{\bf U}}}
\newcommand{\Vmat}[0]{{{\bf V}}}
\newcommand{\Wmat}[0]{{{\bf W}}}
\newcommand{\Xmat}{{\bf X}}
\newcommand{\Ymat}[0]{{{\bf Y}}}
\newcommand{\Zmat}{{\bf Z}}

\newcommand{\av}{\boldsymbol{a}}
\newcommand{\Av}{\boldsymbol{A}}
\newcommand{\Cv}{\boldsymbol{C}}
\newcommand{\bv}{\boldsymbol{b}}
\newcommand{\cv}{{\boldsymbol{c}}}
\newcommand{\dv}{\boldsymbol{d}}
\newcommand{\ev}[0]{{\boldsymbol{e}}}
\newcommand{\fv}{\boldsymbol{f}}
\newcommand{\Fv}[0]{{\boldsymbol{F}}}
\newcommand{\gv}[0]{{\boldsymbol{g}}}
\newcommand{\hv}[0]{{\boldsymbol{h}}}
\newcommand{\iv}[0]{{\boldsymbol{i}}}
\newcommand{\jv}[0]{{\boldsymbol{j}}}
\newcommand{\kv}[0]{{\boldsymbol{k}}}
\newcommand{\lv}[0]{{\boldsymbol{l}}}
\newcommand{\mv}[0]{{\boldsymbol{m}}}
\newcommand{\nv}{\boldsymbol{n}}
\newcommand{\ov}[0]{{\boldsymbol{o}}}
\newcommand{\pv}[0]{{\boldsymbol{p}}}
\newcommand{\qv}[0]{{\boldsymbol{q}}}
\newcommand{\rv}[0]{{\boldsymbol{r}}}
\newcommand{\sv}[0]{{\boldsymbol{s}}}
\newcommand{\tv}[0]{{\boldsymbol{t}}}
\newcommand{\uv}[0]{{\boldsymbol{u}}}
\newcommand{\vv}{\boldsymbol{v}}
\newcommand{\wv}{\boldsymbol{w}}
\newcommand{\Wv}{\boldsymbol{W}}
\newcommand{\xv}{\boldsymbol{x}}
\newcommand{\yv}{\boldsymbol{y}}
\newcommand{\Xv}{\boldsymbol{X}}
\newcommand{\Yv}{\boldsymbol{Y}}
\newcommand{\zv}{\boldsymbol{z}}

\newcommand{\Gammamat}[0]{{\boldsymbol{\Gamma}}}
\newcommand{\Deltamat}[0]{{\boldsymbol{\Delta}}}
\newcommand{\Thetamat}{\boldsymbol{\Theta}}
\newcommand{\Lambdamat}{{\boldsymbol{\Lambda}}}
\newcommand{\Ximat}[0]{{\boldsymbol{\Xi}}}
\newcommand{\Pimat}[0]{{\boldsymbol{\Pi}} }
\newcommand{\Sigmamat}{\boldsymbol{\Sigma}}
\newcommand{\Upsilonmat}[0]{{\boldsymbol{\Upsilon}} }
\newcommand{\Phimat}{\boldsymbol{\Phi}}
\newcommand{\Psimat}{\boldsymbol{\Psi}}
\newcommand{\Omegamat}{{\boldsymbol{\Omega}}}

\newcommand{\Lambdav}{\bm{\Lambda}}
\newcommand{\alphav}{\boldsymbol{\alpha}}
\newcommand{\betav}[0]{{\boldsymbol{\beta}} }
\newcommand{\gammav}{{\boldsymbol{\gamma}}}
\newcommand{\deltav}[0]{{\boldsymbol{\delta}} }
\newcommand{\epsilonv}{\boldsymbol{\epsilon}}
\newcommand{\zetav}[0]{{\boldsymbol{\zeta}} }
\newcommand{\etav}[0]{{\boldsymbol{\eta}} }
\newcommand{\thetav}{\boldsymbol{\theta}}
\newcommand{\iotav}[0]{{\boldsymbol{\iota}} }
\newcommand{\kappav}{{\boldsymbol{\kappa}}}
\newcommand{\lambdav}[0]{{\boldsymbol{\lambda}} }
\newcommand{\muv}{\boldsymbol{\mu}}
\newcommand{\nuv}{{\boldsymbol{\nu}}}
\newcommand{\xiv}{{\boldsymbol{\xi}}}
\newcommand{\omicronv}[0]{{\boldsymbol{\omicron}} }
\newcommand{\piv}{\boldsymbol{\pi}}
\newcommand{\rhov}[0]{{\boldsymbol{\rho}} }
\newcommand{\sigmav}[0]{{\boldsymbol{\sigma}} }
\newcommand{\tauv}[0]{{\boldsymbol{\tau}} }
\newcommand{\upsilonv}[0]{{\boldsymbol{\upsilon}} }
\newcommand{\phiv}{\boldsymbol{\phi}}
\newcommand{\chiv}[0]{{\boldsymbol{\chi}} }
\newcommand{\psiv}{\boldsymbol{\psi}}
\newcommand{\omegav}[0]{{\boldsymbol{\omega}} }

\newcommand{\xin}[1]{{\textcolor{red}{#1}}}

\newcommand{\ts}{^{\top}}
\newcommand{\TV}{{\rm TV}}
\newtheorem{definition}{Definition}
\newtheorem{lemma}{Lemma}
\newtheorem{corollary}{Corollary}
\newtheorem{theorem}{Theorem}

\twocolumn[{%
\renewcommand\twocolumn[1][]{#1}%
\maketitle
\begin{center}
\vspace{-10pt}
    \includegraphics[width=1.\linewidth]{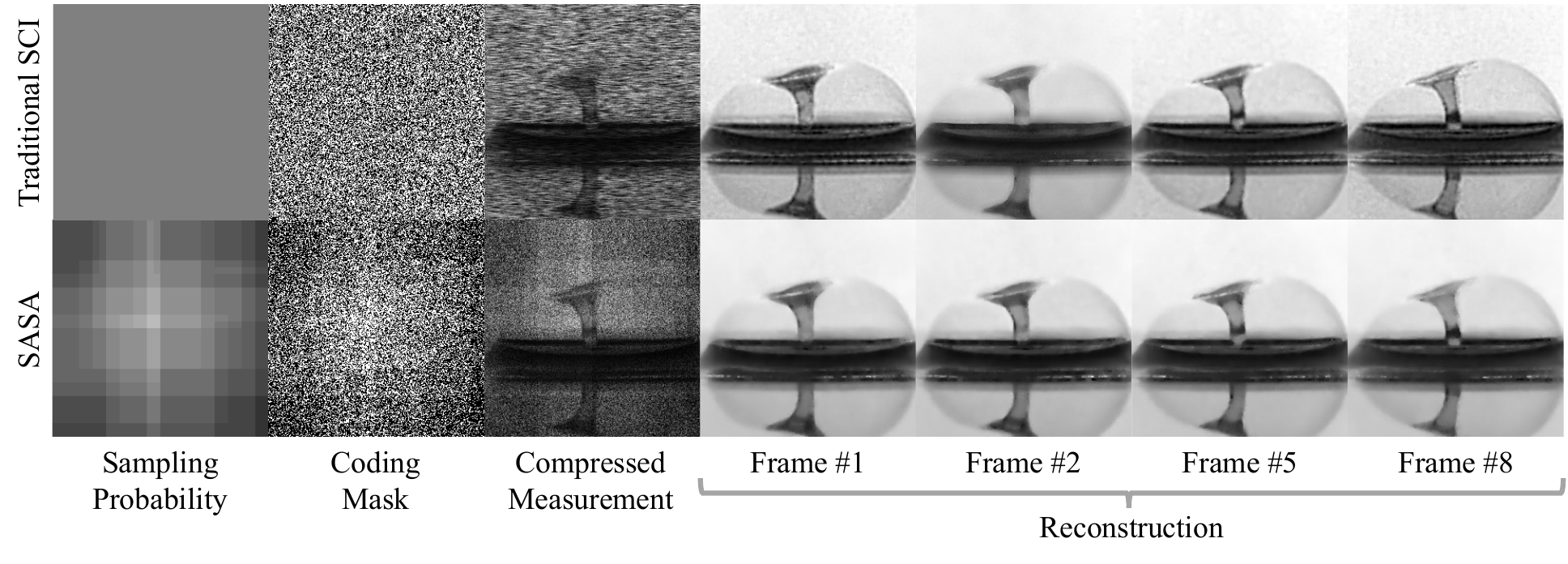}
	\captionof{figure}{
While traditional snapshot compressive imaging (SCI) directly generates random binary matrices as coding masks to obtain compressed measurements, we propose a framework elegantly with saliency detection to perform self-adaptive sampling, and thus boosting the quality of compressed measurement and the performance of reconstruction result.
}
	\label{fig:teaser}
	\vspace{10pt}
\end{center}
}
]

\newcommand\blfootnote[1]{%
  \begingroup
  \renewcommand\thefootnote{}\footnote{#1}%
  \addtocounter{footnote}{-1}%
  \endgroup
}

\blfootnote{$^{*}$Corresponding author.}
\blfootnote{This work is supported in part by the Research Grants Council (GRF 17201822), by the Research Postgraduate Student Innovation Award (The University of Hong Kong), and by ACCESS –- AI Chip Center for Emerging Smart Systems, Hong Kong SAR.}


\vspace{-5mm}
\begin{abstract}
The ability of snapshot compressive imaging (SCI) systems to efficiently capture high-dimensional (HD) data depends on the advent of novel optical designs to sample the HD data as two-dimensional (2D) compressed measurements. Nonetheless, the traditional SCI scheme is fundamentally limited, due to the complete disregard for high-level information in the sampling process. To tackle this issue, in this paper,
we pave the first mile toward the advanced design of adaptive coding masks for SCI. Specifically, we propose an \textit{efficient} and \textit{effective} algorithm to generate coding masks with the assistance of saliency detection, in a \textit{low-cost} and \textit{low-power} fashion.
Experiments demonstrate the effectiveness and efficiency of our approach. 
Code is available at: \href{https://github.com/IndigoPurple/SASA}{\textcolor{blue}{https://github.com/IndigoPurple/SASA}}.
\end{abstract}


\begin{table*}[]
    \centering
    \begin{tabular}{l|l}
        \toprule
        Notation & Description \\
        \midrule
            \multirow{3}{*}{\centering $\Xmat \in \mathbb{R}^{H\times W\times C}  $ } & The video frames we aim to compress and reconstruct.\\
            &$\forall c =1,\dots, C$, $\Xmat_c = \Xmat(:,:,c) \in \mathbb{R}^{H\times W} $ is the c-th video frame.\\
       & Here, $H$, $W$ and $C$ are the image height, image width and compression rate, respectively.\\
       \midrule
        $\Emat \in \mathbb{R}^{H\times W}$ & The measurement noise. \\
        \midrule
        $\Ymat \in \mathbb{R}^{H\times W}$ & The compressed measurement.\\
        \midrule
        \multirow{3}{*}{\centering $\Amat \in \{0, 1\}^{H\times W\times C}$} & The sensing matrix we use to compress the video frames.\\
        & $\forall h = 1, \dots, H$, $\forall w = 1, \dots, W$, $\forall c = 1, \dots, C$, $\Amat_c = \Amat(:,:,c) \in \{0, 1\}^{H\times W}$ is the c-th coding mask,\\
        & $a_{hwc} = \Amat(h,w,c) \in \{0, 1\} $ is the element on the h-th row and w-th column of the c-th coding mask.\\
        \midrule
    \multirow{2}{*}{\centering $Bernoulli(p)$ } & The Bernoulli distribution, 
    which is the discrete probability distribution of a random variable which\\
    & takes the value 1 with probability $p$ and the value 0 with probability $q=1-p$.
    \\
    \midrule
        \multirow{3}{*}{\centering $\Smat \in \mathbb{R}^{H\times W \times D}$} & The saliency maps, where $\Smat_{d} =  \Smat(:,:,d) \in \{0, 1\}^{H\times W}$, is the d-th saliency map.\\
        & Here, value 1 indicates saliency, and value 0 indicates no saliency.\\
        & $D$ is the maximum number of detections to examine. \\
        \midrule
        $\Pmat \in \mathbb{R}^{H\times W}$ & The sampling probability, where $p_{hw} =  \Pmat(h,w) \in [0, 1]$ is a probability parameter.\\
         \bottomrule
    \end{tabular}
    \caption{Key notations and descriptions.}
    \label{tab:notation}
\end{table*}

\section{introduction}
\label{sec:intro}

Snapshot compressive imaging (SCI) systems address the challenges associated with high-dimensional (HD) visual signal acquisition and analysis~\cite{yuan2021snapshot}. These systems capture HD data as compressed two-dimensional (2D) measurements, and then utilize compressed sensing algorithms to reconstruct the HD data~\cite{chan2016plug,yuan2016generalized,zhao2023deep,yang2022revisit}.
By employing optical designs, SCI systems enable the acquisition of HD visual signals in a single snapshot, as opposed to conventional methods. This capability proves advantages in scenarios where capturing dynamic or fast-moving scenes is crucial. Therefore, SCI systems offer promising solutions for real-world applications in fields such as surveillance, biomedical imaging, remote sensing

The key principle behind SCI lies in the compression of the acquired HD data. Rather than directly sampling the entire HD signal, SCI systems capture a compressed version of the data, resulting in compressed measurements. 
However, the existing research in this area has predominantly focused on the reconstruction algorithms, while the study of compression sampling methods remains lacking, let alone optimizing the mask design. Additionally, the current sampling methods treat all pixels equally by employing random binary masks, completely disregarding high-level information such as objects and saliency. As a result, the importance of different regions of the image is not properly considered.

An even more serious issue is that the \textit{``sampling - reconstruction - tasks"} scheme of traditional SCI is fundamentally limited. On one hand, the complete ignorance of high-level information in the sampling process leads to wasteful resource allocation for content-irrelevant computation. On another hand, the independence of subsequent computer vision tasks hampers the sampling and design efficiency of SCI~\cite{zhang2022compressive}.

To solve those problems, as Figure~\ref{fig:teaser} shows, we propose an \textit{efficient} and \textit{effective} algorithm to generate adaptive coding masks with the assistance of saliency detection, in a \textit{low-cost} and \textit{low-power} fashion. Our key insight lies in that the sampling process to acquire compressed measurements should be performed adaptively and intelligently, which assigns a higher sampling probability to salient image regions and a lower probability to non-salient regions.

Our main contributions are as follows:
\begin{itemize}
    \item To the best of our knowledge, this work is \textbf{the first} to design adaptive coding masks specifically for SCI.
    \item Our proposed method introduces a \textbf{low-cost and efficient} approach by performing saliency detection directly on the measurement. This strategy requires only a few atomic operations, making it highly suitable for real-time applications. Notably, the processing speed is on average 250fps on a single laptop CPU.
    \item Quatitative and qualitative experiments on classical datasets demonstrate the \textbf{effectiveness} of our method.
\end{itemize}

\begin{figure*}
    \centering
    \includegraphics[width =\linewidth]{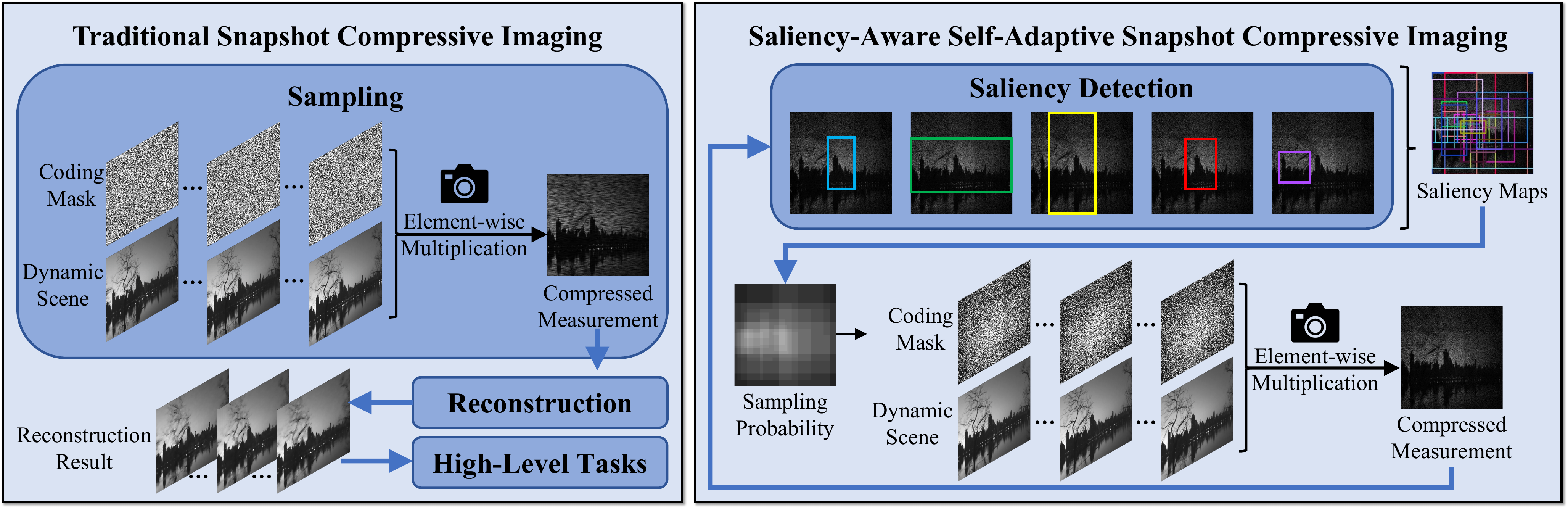}
    \caption{The pipeline of traditional SCI and our proposed saliency-aware self-adaptive SCI, respectively.}
    \label{fig:pipeline}
\end{figure*}

\section{related work}


SCI systems consist of two components: 
an encoding process that compresses the HD data into a 2D measurement by hardware~\cite{wagadarikar2008single,llull2013coded,yuan2021snapshot}, and a decoding process that reconstructs the desired signal by algorithms~\cite{yuan2016generalized,yuan2020plug,yang2022revisit,liu2018rank,Yang14GMMonline,Yang14GMM,yuan2016generalized,qiao2020deep,zheng2021super, wang2022spatial, cheng2022recurrent, meng2021perception,meng2020gap,wu2021dense,zhao2022mathematical,zhao2023transmission,zhenyuen2023solving}. Nevertheless,
the existing research in this area focused
on the reconstruction algorithms, leaving a gap in the study of compression sampling methods.

Despite recent efforts to bridge the gap between compressive imaging and computer vision, such as directly performing vision tasks on the compressed measurements rather than the reconstructed data~\cite{lu2020edge, zhang2022compressive}, the fundamental issue remains unresolved: the complete disregard of high-level information during the sampling process. As a result, there is a pressing need to conduct research on optimizing the mask design. Traditional SCI systems~\cite{yuan2021snapshot} treat all pixels equally, employing random binary masks that fail to consider crucial high-level information, such as objects and saliency. Consequently, the importance of different regions within an image is not appropriately taken into account.

Therefore, in this paper, we aim to pave the first mile towards an adaptive SCI system that leverages high-level information during the sampling process.

\section{method}
\label{sec:method}

In this section, we first formulate the sampling process in video SCI. Next,  we propose an \textit{efficient} and \textit{effective} algorithm to generate adaptive coding masks, in a \textit{low-cost} and \textit{low-power} fashion. Table~\ref{tab:notation} outlines key notations used in this paper.

Figure~\ref{fig:pipeline} provides an intuitive understanding of our pipeline, jointly with the comparison to traditional SCI. While traditional SCI directly generates random binary matrices as coding masks to obtain compressed measurements, our framework elegantly utilizes saliency detection to perform self-adaptive sampling.

To begin with, the acquisition of compressed measurements in video SCI could be mathematically modeled as:
\begin{align}
    \Ymat = \sum^C_{c=1}\Amat_c\odot\Xmat_c + \Emat,
\end{align}
where $\Ymat$ is the compressed measurement, $\Amat$ is the sensing matrix, $\Xmat$ is the video frames we aim to compress and reconstruct, $\Emat$ is the measurement noise, and $\odot$ denotes the Hadamard (element-wise) product.

In traditional SCI~\cite{yuan2021snapshot}, the sensing matrix $\Amat$ is always generated by randomly sampling elements from a Bernoulli distribution with probability $p=0.5$. In our SCI system, we initialize the first sensing matrix by the traditional method:
\begin{align}
\begin{aligned}
\Amat^{(0)} = \big[a_{hwc}^{(0)}\big],\qquad a_{hwc}^{(0)} \sim Bernoulli(0.5),\\
    \forall h = 1, \dots, H,\ \forall w = 1, \dots, W,\ \forall c = 1, \dots, C,
\end{aligned}
\end{align}
which is used to preliminary compress the dynamic scene and capture the first measurement:
\begin{align}
    \Ymat^{(0)} = \sum^C_{c=1}\Amat^{(0)}_c\odot\Xmat^{(0)}_c + \Emat^{(0)}.
\end{align}
Once the first measurement is obtained, we perform advanced vision processing technology, \textit{i.e.}, saliency detection directly on the measurement in stead of on the reconstructed video, to boost inference speed and reduce bandwidth occupation, memory footprint and energy consumption. Specifically, we adopt the light-weight algorithm from~\cite{cheng2014bing}, denoted as $f$, which takes the measurement $\Ymat^{(t)}$ corresponding to the $t$-th video sequence as input and outputs saliency maps $\Smat^{(t)}$:
\begin{align}
    \Smat^{(t)} = f(\Ymat^{(t)}; D),\qquad t = 0,1,2,\dots,
\end{align}
where $D$ is a parameter that determines the maximum number of detections to examine. Based on the estimated saliency maps, we calculate the sampling probability for the next video sequence by retrieving saliency in the compressed domain with low bandwidth:
\begin{align}
    \Pmat^{(t+1)} =  \frac{1}{D}\sum^D_{d=1} \Smat_d^{(t)},\qquad t = 0,1,2,\dots,
\end{align}
where $\Pmat$ is the sampling probability used to guide sensing matrix generation for the next video sequence. It assigns higher sampling probabilities to salient image regions and lower to non-salient, ensuring an adaptive and effective sampling process. To ensure sampling coverage in regions without salient events, we then assign a probability of $1/D$ to areas with zero probability. According to $\Pmat = [p_{hw}]$, the sensing matrix is updated by:
\begin{align}
\begin{aligned}
    \Amat^{(t+1)} = \big[ a^{(t+1)}_{hwc} \big],\qquad a^{(t+1)}_{hwc}\sim Bernoulli\big(p_{hw}^{(t+1)}\big),\\
    \forall h = 1, \dots, H,\ \forall w = 1, \dots, W,\ \forall c = 1, \dots, C,\\
    t = 0,1,2,\dots.
\end{aligned}
\end{align}
Here, our key insight lies in that the sampling process of acquiring compressed measurements should be performed adaptively, which assigns a higher sampling probability to salient image regions and lower to non-salient regions.

Finally, the measurement of the next video sequence is captured using the dynamically updated sensing matrix:
\begin{align}
\begin{aligned}
    \Ymat^{(t+1)} = \sum^C_{c=1}\Amat^{(t+1)}_c\odot\Xmat^{(t+1)}_c + \Emat^{(t+1)},\\ t= 0, 1, 2,\dots.
\end{aligned}
\end{align}


It is worth noting that our method performs saliency detection directly on the measurement, and requires only a few atomic operations to generate sensing matrices. In our experiments, the processing speed is on average 250fps on a single laptop CPU, which means that \textbf{it enables a low-speed camera to capture high-speed scenes with $C\times250$ fps} while intelligently and adaptively varying the coding masks.


\begin{figure}
    \centering
    \includegraphics[width =\linewidth]{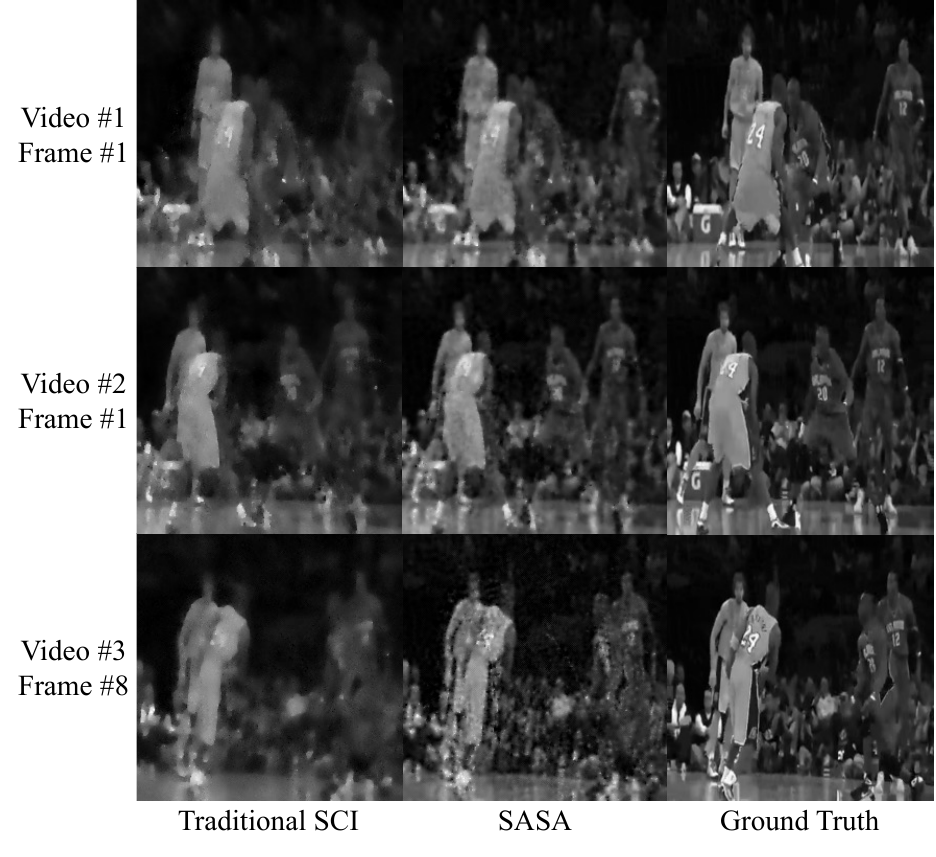}
    \caption{Comparisons of reconstruction results on the dataset \texttt{Kobe}. Reconstruction is by ADMM-TV~\cite{chan2016plug}.
 }
    \label{fig:kobe}
\end{figure}

\begin{figure}
    \centering
    \includegraphics[width =\linewidth]{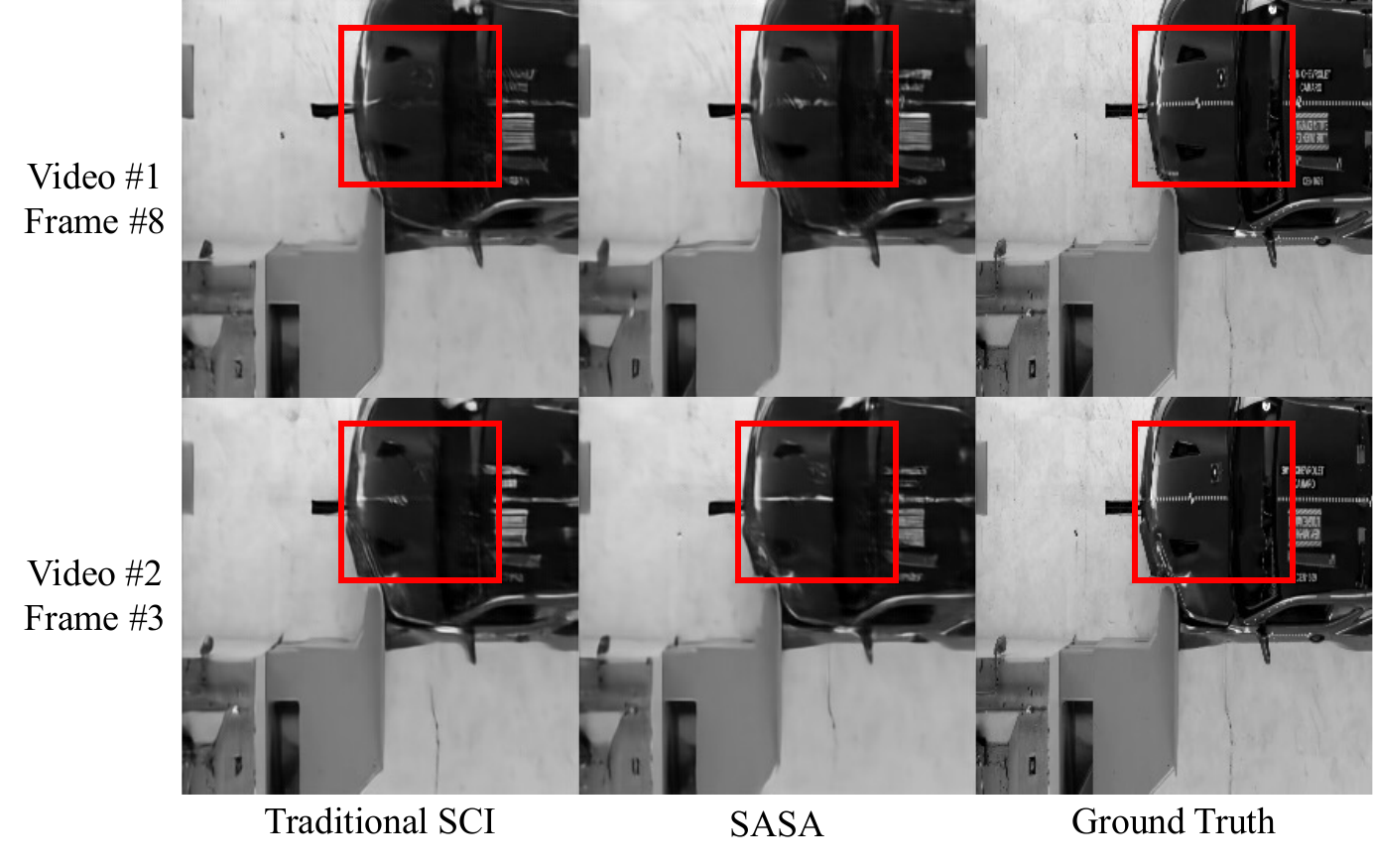}
    \caption{Comparisons of reconstruction results on the dataset \texttt{Vehicle}. Reconstruction is by DEQSCI~\cite{zhao2023deep}.
 }
    \label{fig:vehicle}
\end{figure}

\begin{table}[]
    \centering
    \resizebox{1\linewidth}{!}{
    \begin{tabular}{c|cccc|c}
    \toprule
        Dataset & Aerial & Vehicle & Kobe & Traffic & Average \\
        \midrule
        Recon. & \multicolumn{5}{c}{ADMM-TV~\cite{chan2016plug}}
        \\
        \midrule
        Trad. SCI & 24.54, 0.877 & 23.88, 0.822 & 25.38, 0.821 & 20.15, 0.740 & 23.49, 0.815\\
        SASA & \textbf{24.61, 0.882} & \textbf{24.44, 0.862} & \textbf{26.35, 0.877} & \textbf{20.19, 0.746} & \textbf{23.90, 0.842}\\
        \midrule
        Recon. & \multicolumn{5}{c}{GAP-TV~\cite{yuan2016generalized}}
        \\
        \midrule
        Trad. SCI & 24.69, 0.861 & 24.51, 0.867 & 25.97, 0.863 & 20.44, 0.763 & 23.90, 0.839\\
        SASA & \textbf{24.71, 0.864} & \textbf{24.57, 0.872} & \textbf{26.80, 0.895} & \textbf{20.48, 0.771} & \textbf{24.13, 0.851}\\
        \bottomrule
    \end{tabular}
    }
    \caption{The results in terms of PSNR (dB) and SSIM by different compressed measurements on classical four datasets. 
    }
    \label{tab:exp}
\end{table}

\begin{table}[]
    \centering
    \resizebox{1\linewidth}{!}{
    \begin{tabular}{c|cccc|c}
    \toprule
        $D$ & Aerial & Vehicle & Kobe & Traffic & Average \\
        
        \midrule
        10 & 24.77, 0.871 & 23.69, 0.867 & 25.16, 0.845 & 19.92, 0.740 & 23.39, 0.831\\
        20 & 24.55, 0.859 & 24.13, 0.856 & 25.89, 0.857 & 20.39, 0.759 & 23.74, 0.833\\
        30 & \textbf{24.71, 0.864} & \textbf{24.57, 0.871} & \textbf{26.80, 0.895} & \textbf{20.48, 0.770} & \textbf{24.13, 0.850}\\
        40 & 24.56, 0.865 & 24.53, 0.873 & 26.62, 0.896 & 20.26, 0.754 & 23.99, 0.847\\
        50 & 24.42, 0.862 & 24.41, 0.871 & 26.48, 0.896 & 20.16, 0.745 & 23.87, 0.843\\
        \bottomrule
    \end{tabular}
    }
    \caption{Ablation study on different max detection numbers in saliency detection. Reconstruction is by GAP-TV~\cite{yuan2016generalized}.}
    \label{tab:as}
\end{table}

\section{experiment}
\label{sec:exp}
\vspace{-10pt}
\textbf{Experimental Setting. } We initialize the first sensing matrix and compressed measurement following the traditional SCI~\cite{yuan2021snapshot}, and then use the proposed method to generate adaptive sensing matrices for capturing sequential compressed measurements. For evaluations, four of the six classical SCI datasets~\cite{yuan2020plug} including  \texttt{Aerial}, \texttt{Vehicle}, \texttt{Kobe} and \texttt{Traffic} are adopted, while two (\texttt{Drop} and \texttt{Runner}) are excluded because the videos are too short and they contain only a single measurement. Then two classical SCI reconstruction algorithms ADMM-TV~\cite{chan2016plug} and GAP-TV\cite{yuan2016generalized}, and a state-of-the-art method DEQSCI~\cite{zhao2023deep,zhao2022deep} are conducted, on different measurements captured by traditional SCI and our proposed framework (SASA), to reconstruct the videos.

\noindent{\textbf{Quantitative and Qualitative Evaluations. }} Quantitative comparison results of different compressed measurements are provided in Table~\ref{tab:exp}, where our method achieves approximately 0.2$\sim$0.5 dB improvement in PSNR and 0.01$\sim$0.03 in SSIM~\cite{wang2004image} on average. The improvement indicates our method could reconstruct images with relatively fine structure, which is confirmed by qualitative evaluations in Figures~\ref{fig:kobe} and~\ref{fig:vehicle}. As shown in these figures, reconstruction results from traditional SCI have more artifacts and distortions around margins. Our method can maintain a clear and accurate image structure, thus leading to higher performance. 
Ours adjusts the coding masks during the sampling process, seamlessly integrates with existing reconstruction algorithms, and enhances performance across all baselines. Furthermore, the reconstruction quality can be further improved using various video processing techniques~\cite{zhao2021efenet,zhao2022manet,zhao2021cross,zhao2022cross,zhao2023improving}.

\noindent{\textbf{Ablation Study. }} In Table~\ref{tab:as}, we experimented with different values for $D$, the maximum number of detections to examine in saliency detection, and found $D=30$ is the best setting.

\section{conclusion}
This work addresses the limitations of existing SCI systems by an efficient and effective framework to adjust coding masks, paving the first mile towards adaptive SCI systems. 

\small
\bibliographystyle{IEEEtran}
\bibliography{refs_sasa}

\end{document}